\documentclass[english]{article}
\usepackage[T1]{fontenc}
\usepackage[utf8]{luainputenc}
\usepackage{geometry}
\usepackage[active]{srcltx}
\usepackage{float}
\usepackage{amsbsy}
\usepackage{amstext}
\usepackage{amssymb}
\usepackage{graphicx}
\usepackage{hyperref}
\usepackage{setspace}
\usepackage{amsmath} 

\newcommand{\footremember}[2]{%
    \footnote{#2}
    \newcounter{#1}
    \setcounter{#1}{\value{footnote}}%
}
\newcommand{\footrecall}[1]{%
    \footnotemark[\value{#1}]%
} 
\makeatletter

\providecommand{\tabularnewline}{\\}

\makeatother

\usepackage{babel}

\title{Bias extension test on an unbalanced woven composite reinforcement: Experiments and modeling via a second-gradient continuum approach}
\author{
Gabriele Barbagallo\footremember{LaMCos}{LaMCoS, Université de Lyon, INSA-CNRS, 20 avenue Albert Einstein, 69621,Villeurbanne cedex}\footremember{LGCIE}{LGCIE, Université de Lyon, INSA, 20 avenue Albert Einstein, 69621, Villeurbanne cedex}, Angela Madeo\footrecall{LGCIE} \footremember{MEMOCS}{International research center M\&MoCS, Università dell'Aquila, Palazzo Caetani, 04012 Cisterna di Latina (LT), Italy}\footremember{Cor}{Corresponding author, e-mail:angela.madeo@insa-lyon.fr}, Ismael Azehaf\footrecall{LaMCos}, Ivan Giorgio\footrecall{MEMOCS} \footremember{Sapienza}{Dipartimento di Meccanica e Aeronautica, Università Roma Sapienza, via Eudossiana 18,00184 Roma, Italy},\and Fabrice Morestin\footrecall{LaMCos} and Philippe Boisse\footrecall{LaMCos}}
\date{}

\begin{document}
\maketitle

\begin{abstract}
The classical continuum models used for the woven fabrics do not fully describe the whole set of phenomena that occur during the testing of those materials. This incompleteness is partially due to the absence of energy terms related to some micro-structural properties of the fabric and, in particular, to the bending stiffness of the yarns. To account for the most fundamental microstructure-related deformation mechanisms occurring in unbalanced interlocks, a second-gradient, hyperelastic, initially orthotropic continuum model is proposed. 

A constitutive expression for the strain energy density is introduced to account for i) in-plane shear deformations, ii) highly different bending stiffnesses in the warp and weft directions and iii) fictive elongations in the warp and weft directions which eventually describe the relative sliding of the yarns. Numerical simulations which are able to reproduce the experimental behavior of unbalanced carbon interlocks subjected to a Bias Extension Test are presented. In particular, the proposed model captures the macroscopic asymmetric S-shaped deformation of the specimen, as well as the main features of the associated deformation patterns of the yarns at the mesoscopic scale. 
\end{abstract}

\section*{Introduction}
The use of textile composites has allowed the conception of engineering pieces with improved performances, but there is still a need for the development of new theories and softwares for the accurate modeling of the mechanical response of such materials during the forming processes of industrial products and components. The internal micro-structure of the woven fabrics' reinforcements strongly influences their mechanical properties and makes models capable of capturing all the aspects of their complex mechanical behavior challenging to set up.

Two main directions of woven tows (warp and weft) with each composed by a high number of fibers define the structure of the woven fabric. Because of the preferred directions generated by the fiber lines, the material possesses two usually orthogonal directions with a very high extensional rigidity. Among the various industrial and commercial woven composites, it is possible to highlight various schemes of weaving which confer different mechanical properties to the fabric. Moreover, another characteristic that can profoundly determine the response of the material is the ratio between the weights (or sizes) of the yarns in the warp and weft directions. If the ratio is not equal to one, the woven fabric is called ``unbalanced'', otherwise it is called ``balanced''. In an unbalanced fabric, the properties in the two weaving directions strongly differ giving rise to a stiffer and/or stronger direction that can be of use for particular engineering applications. For example, if there is one main direction of loading, the use of an unbalanced fabric allows for the optimization of the material bringing real design advantages. 

During the process of weaving, small yarns or tows are woven to form a complex texture. These yarns are made with materials that possess high specific mechanical properties such as the traditional carbon and glass fibers or even polymeric and ceramic fibers. Furthermore, in the case of the carbon fiber reinforcements, any single yarn is, itself, composed of thousands of small carbon fibers. This complex hierarchical microstructure characterizes the global features of the material. In a wealth of real cases, it is reasonable to consider that the friction between the yarns prevents the slipping and contributes to the shear rigidity of the fabric that fundamentally determines its mechanical behavior. Nevertheless, to comprehensively describe the behavior of woven fabrics, other phenomena besides the shear stiffness of the fabric and the elongation stiffness of the yarns, which can be ideally thought to be quasi-inextensible, must be taken into account. When the yarns that compose the fabric are relatively thick, they possess a relevant bending stiffness at the meso-level that accounts for some specific experimental response (see \cite{MadeoBias}) and, in the case of unbalanced fabrics, the difference in the aforementioned bending stiffness can convey some interesting asymmetric macroscopic deformations.

Woven fabrics are materials that possess very interesting features in terms of specific stiffness and strength, deformability, dimensional stability, thermal expansion, corrosion resistance and many other properties, basically due to the intrinsic mechanical properties of the material comprising the fibers. Of all such desirable features, the high shear deformability makes it possible for these materials to develop in various shapes. Anyway, without tools to forecast phenomena such as the onset of wrinkling and slippage that limit the admissible deformation during the stamping operations, it is not possible to fully exploit the huge potential of these materials. Hence, the development of comprehensive models for the description of the forming of these kinds of materials it is of primary importance.

In the past years, researchers have proposed different approaches to such textile forming problems mainly focusing on the development of discrete and continuum models which are able to account for the basic deformation mechanisms that occur during the forming of woven composite reinforcements.

In the first part of this paper, the results of some experiments, that clearly show how the effect of the unbalance of the warp and weft material properties influences the overall macroscopic behavior of the considered composite interlocks, are presented. More particularly, it will be shown that, from an experimental point of view, unbalanced fabrics with very different local bending stiffness give rise to asymmetric macroscopic deformations during a bias-extension test. In fact, different complex phenomena take place at the mesoscopic level which are, to a big extent, related to 
\begin{itemize}
\item the in plane shear deformation (change of direction of the warp and weft yarns),
\item the high differential bending rigidity of the two families of yarns,
\item the relative sliding of the two families of yarns.
\end{itemize}
In particular, the strong unbalance in the bending rigidities results in an S-shaped deformed specimen. 

A second-gradient continuum model, able to capture the basic features of those phenomena, is introduced and some numerical simulations, which allow interpretation of the results of the experiments presented in this paper, are proposed. The main scope of the present paper is to propose a continuum model which is able to account for all the aforementioned microstructure-related deformation mechanisms by remaining in a continuum framework. It will be shown that 
\begin{itemize}
\item the shear deformation can be included by introducing a suitable energetic cost associated with the angle variation between the warp and weft directions, 
\item the differential bending rigidities of the two families of yarns can be considerd by means of the introduction of second-gradient terms and such unbalance is responsible for the asymmetric S-shape of the specimen,
\item the relative sliding of the fibers can be included in the continuum model by introducing ``fictive'' elongations in the warp and weft directions.
\end{itemize}

In what follows, it will be denoted by ``fictive'' elongation, an elongation of the yarns as accounted for by the introduced continuum model. The adjective ``fictive''  wants to stress the fact that such elongation in the continuum model actually represents, to a big extent, a yarns' sliding in the real mechanical system. This expedient allows to keep using a continuum model, even if the micro-displacement field of the real system suffers tangent jumps due to the relative motion of the yarns.

Complex theoretical formulations of the considered problem are not given, the main aim being that of noting some peculiar behaviors of unbalanced fabrics based on phenomenological observations. The introduced equivalent continuum model can be seen as a reasonable compromise between the detail of the description of the behavior of the underlying microstructures and the complexity of the adopted model.

\section{The Bias Extension Test}

To characterize the macroscopic mechanical response of the woven composite reinforcements, two set of experimental tests are proposed. Indeed, the behavior of such materials must be analyzed under different types of loads and, in particular, the most important feature to be determined is the in-plane shear response of the woven composite. In fact, due to the quasi-inextensibility of the fibers, the main deformation mode of the woven composites reinforcements during a forming process is the in-plane shear deformation (angle variation between the warp and weft). 

Two main tests are of current use for the determination of the in-plane shear stiffness of the fabrics. The first test developed was the Picture Frame Test (PFT). In the PFT, a square specimen of the woven composite is ideally subjected to a state of pure shear deformation. Nevertheless, the state of pure shear is only theoretical:  any misalignment of the specimen leads to an increase of the measured load \cite{Lomov2006,Launay2008,Nosrat2014}. In addition, the fact that the yarns are tightly clamped fixes the direction of the yarns correspondingly to the four clamps and this fact generates bending of the fibers in the vicinity of the boundary during the motion.  This fact usually results in an overestimation of the shear parameter due to these boundary effects.

\begin{figure}[H]
\begin{centering}
\includegraphics[scale=0.8]{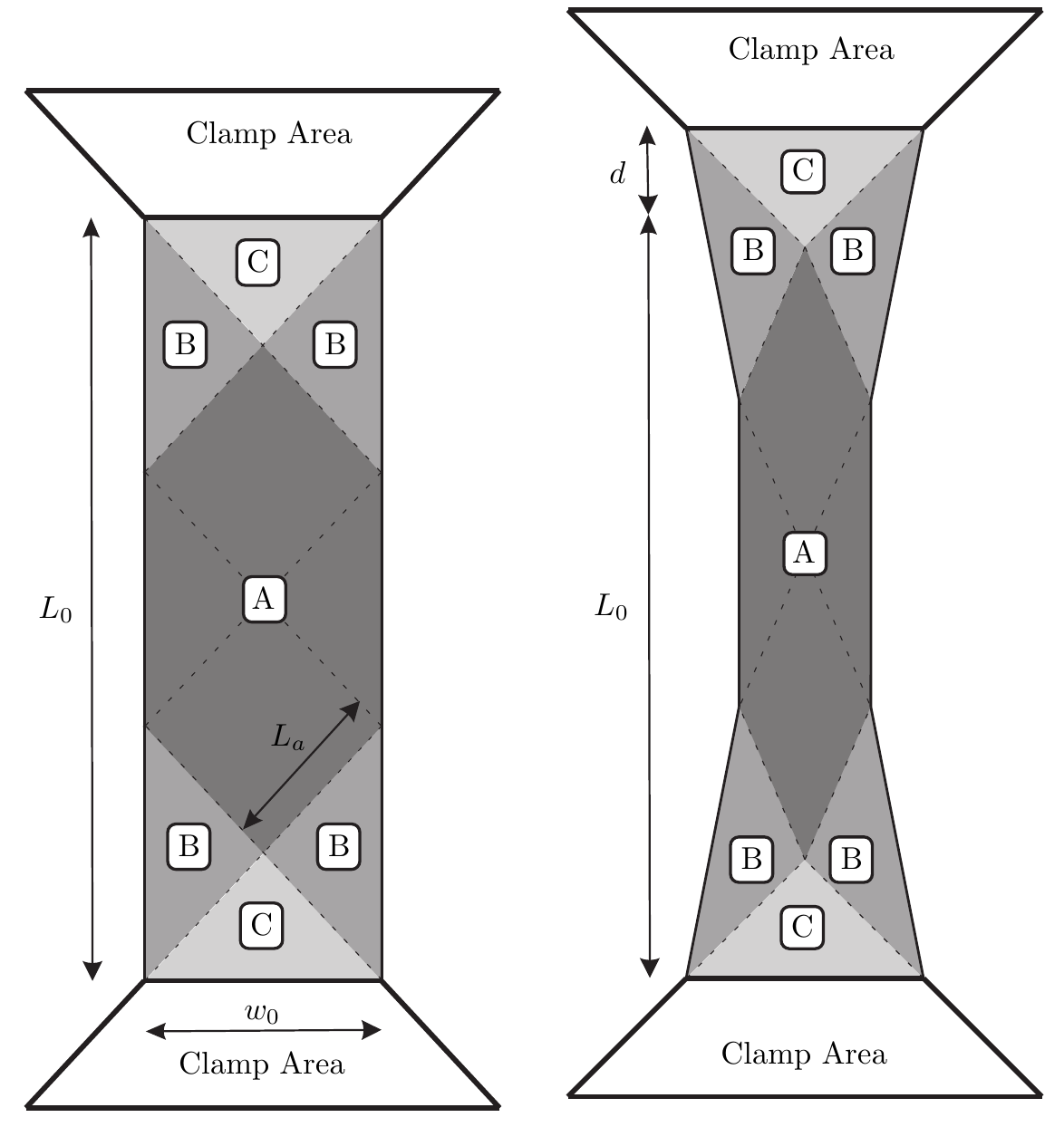}
\par\end{centering}
\protect\caption{Simplified description of the shear angle pattern in the bias extension test\label{fig:BET-Schematics}}
\end{figure}

The second test used for the measure of the in-plane shear stiffness is the bias-extension test (BET). In this test one of the edges of a rectangular sample of woven composite reinforcement, in which the yarns are initially oriented at $\pm45^{\circ}$ with respect to the loading direction, is displaced in the direction of the axis \cite{Lee e Phil,Cao Phi,Harrison2008,Peng2004,Potter2002,Wang1998}. The length/width ratio of the specimen must be larger than 2 (in the present paper the ratio is 3). When one of the ends of the specimen is displaced of a given amount, the formation of three types of regions (A, B and C) with almost homogeneous behavior in their interior can be observed (figure \ref{fig:BET-Schematics}). In each of these areas, the angle between the warp and weft direction is almost constant. This specific kinematics is due to the quasi-inextensibilty of the yarns and to rotation without slippage between warp and weft yarns at the crossover points. The advantage of the BET, with respect to the PFT, is that each yarn has at least one edge which is free, and this free edge is thought to be sufficient to avoid spurious tensions in the yarns as observed in \cite{Harrison,Launay2008}.

\begin{figure}[H]
\begin{centering}
\includegraphics[scale=1.2]{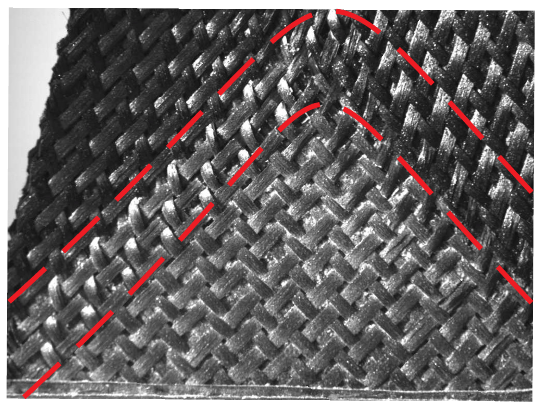}
\par\end{centering}

\protect\caption{Boundary layers between two regions at constant shear\label{fig:BET-Boundary}}
\end{figure}

Still, there are some phenomena that are not described in the schematics of figure \ref{fig:BET-Schematics} and which are nevertheless important for the complete understanding of the BET. As a matter of fact, the transition between two different areas at constant shear angle is not concentrated in a line. In fact, it is possible to observe the onset of transition layers, in which there is a gradual variation of the angle as shown in figure \ref{fig:BET-Boundary}. In such transition layers, the angle variation between the two constant values is achieved by a smooth pattern which is directly associated to a local bending of the yarns.

\begin{figure}[H]
\begin{centering}
\includegraphics[scale=0.4]{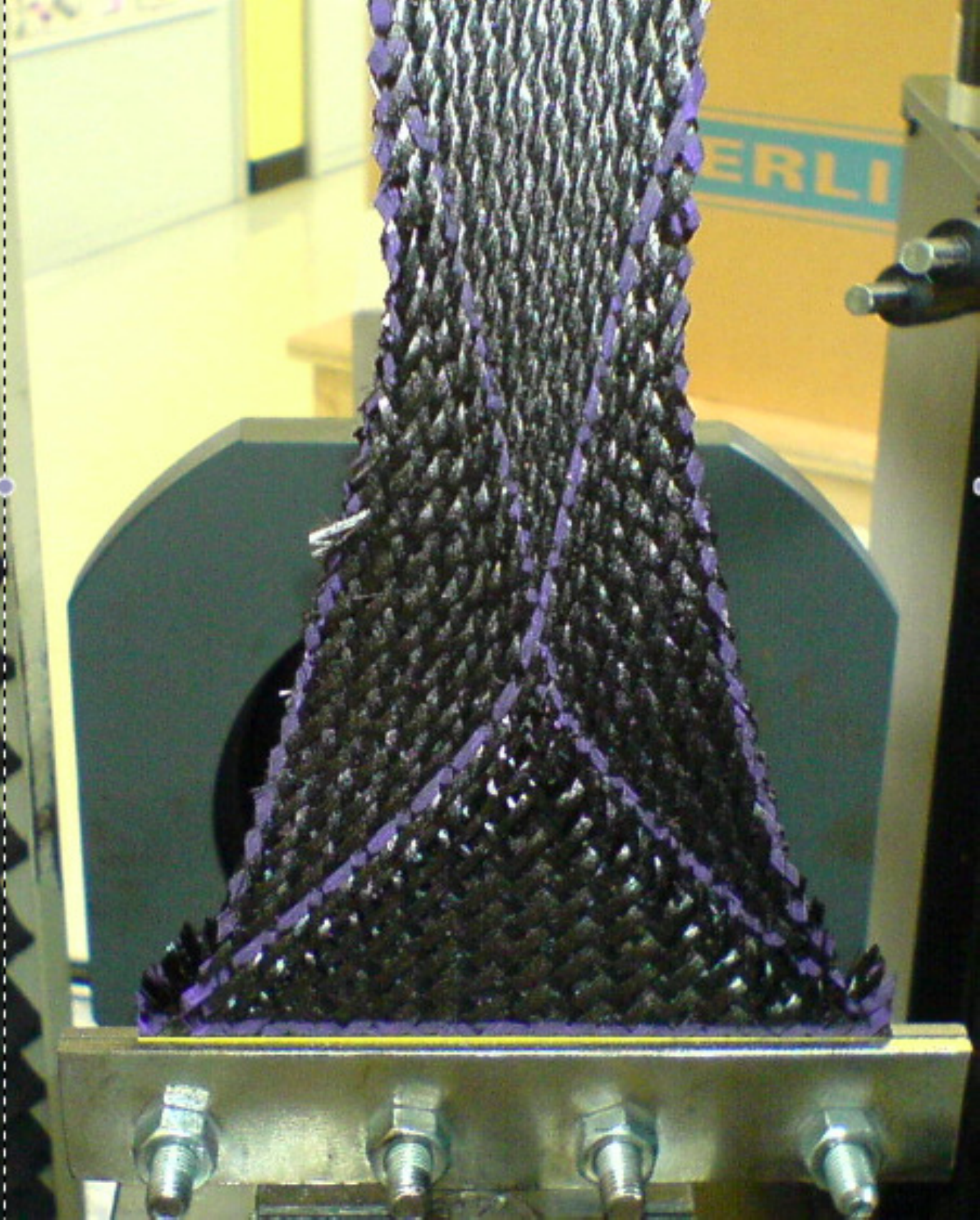}
\par\end{centering}

\protect\caption{Curvature of the free boundary during the BET\label{fig:BET-Curvature}}
\end{figure}

One more feature that can be highlighted is that the free boundary does not remain straight during the test as is assumed in the scheme of figure \ref{fig:BET-Schematics}, but it shows some curvature (figure \ref{fig:BET-Curvature}). Both of these phenomena can be understood by considering that the yarns possess a bending stiffness that depends on the micro-structure of the fabric. Because the yarns always possess a non-vanishing bending stiffness, a variation of orientation cannot be concentrated in a point, but a gradual variation of such orientation takes place (bending of the yarns). Therefore, such angle variations must happen in a layer of non-vanishing size. With the usual first-gradient models, it is not possible to include such micro-structural related phenomena while, with the aid of second-gradient theories, promising results have been obtained in \cite{MadeoBias}. In those models, a specific constitutive coefficient can be introduced which can be directly related to the bending energy of the fibers.

The second-gradient continuum model introduced in the present paper is able to describe the main macroscopic and mesoscopic deformation mecanisms taking place during a bias extension test on an unbalanced fabric. The use of second-gradient theories to capture some effects of the microstructure on the overall behavior of microstructured material is not new and, as a matter of fact, it has been known since the pioneering works by Piola \cite{Piola}, Cosserat \cite{Cosserat}, Midlin \cite{Mindlin}, Toupin \cite{Toupin}, Eringen \cite{EringenBook}, Green and Rivlin \cite{GreenRivlin} and Germain \cite{Germain,Germain2} that many microstructure-related effects in mechanical systems can be still modeled by means of continuum theories. More recently, these generalized continuum theories have been widely developed to describe the mechanical behavior of many complex systems, such as exotic media obtained by homogenization of heterogeneous media \cite{Alibert,Pideri-Sepp,Seppecher Exotic}.

\section{The Bias Extension Test on unbalanced fabrics }

The main aim of the present paper is to show and discuss some specific results obtained during an uniaxial BET of unbalanced specimens of fibrous composite reinforcement. The task of fully exploring what theoretical tools are needed to optimize the modeling of such unbalanced materials is left as a subsequent work, the scope of the present manuscript being that of explaining the principal micro and macro deformation mechanisms which take place when observing the deformation of such unbalanced fabrics. In fact, based on a phenomenological observation of some experimental results, it will be shown that the deformation modes which take place in a BET on an unbalanced fabric are completely different from those specific of the BET on standard fabrics described above. Moreover, a second-gradient continuum model will be proposed for a reasonable description of the micro and macro deformation of such unbalanced materials by simultaneously pointing out the strong and weak points that the employment of such a continuum modeling may have concerning the design of complex engineering parts. 

The authors are aware of the fact that extra experimental campaigns would be needed for a complete validation of the proposed second-gradient model. More particularly, such more comprehensive campaigns would need the setting up of the following experiments:
\begin{itemize}
\item  Repetition of the same test (Bias Extension Test) on different specimens with the same dimensions and characteristics. These tests would be needed to identify experimental errors that can be introduced during the experimental campaign and to precisely  account for such variability in the performed study.
\item Conception of independent tests (other than the Bias Extension Test) which are suitably engineered to give rise to the same microscopic deformation modes (fibers’ bending and slipping), but with different loading conditions. This test would allow the confirmation of the values of the parameters proposed in the present paper for the considered materials.
\item Realization of the Bias Extension Test on specimens with smaller dimensions to unveil possible size effects which are known to be possible in higher gradient or micromorphic materials.
\item Realization of specific measurements which are devoted to measure local deformation mechanisms with the due precision. Digital Image Correlation Techniques could represent a good choice to effectively proceed in this direction.
\end{itemize}
Notwithstanding the undiscussed interest of the aforementioned tests and their necessity for a complete validation of the presented second-gradient model, they are not the primary objective of the present paper. The primary aim is to identify the main microstructure-related deformation modes in unbalanced woven fabrics, i.e.  the differential bending of the fibers and the fibers’ slipping,  and to show, via a reasonable second-gradient model, that they cannot be neglected. The phase of conception of such extra experimental campaigns for a precise identification of second-gradient parameters on a given class of fibrous woven materials is postponed to further investigations.

It has to be explicitly remarked that mechanical conditioning was not accounted for in the present study with the aim of being closer to the conditions of a real forming process.

\subsection{Experimental results}

\begin{figure}[H]
\begin{centering}
\includegraphics[scale=0.5]{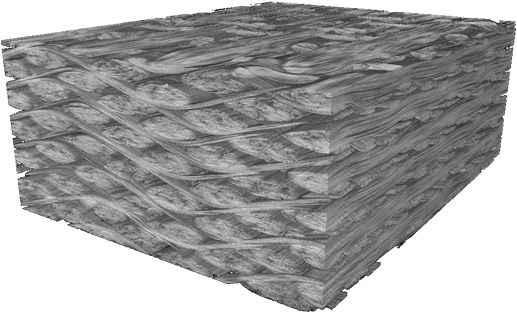}
\par\end{centering}

\protect\caption{\label{fig:TEST-twill}X-ray tomography of the considered interlock
(specimen I).}
\end{figure}

Unbalanced fibrous composite reinforcements are such that the warp and weft yarns are comprised of a very different number of fibers and, therefore, the mechanical properties in the two directions can differ considerably. The material studied in this paper is an unbalanced 2.5 D composite interlock with a characteristic weaving pattern in the direction of the thickness which can be observed in figure \ref{fig:TEST-twill}. The main advantage of interlock reinforcements is to overcome the low delamination fracture toughness of laminated composites \cite{Mouritz2011}. The BET is performed for two samples of 3x3 twill unbalanced carbon interlocks. The two specimens differ one from the other because they present different unbalance ratios between the warp and weft. To analyze the structure of the given specimen, it is possible, using Tomographic Optical Systems, to obtain some virtual 'slices' (Tomographic images) that allow us to see inside the object without cutting. Tomographic Optical Systems is a technology that, through the use of any kind of penetrating wave, reconstructs the geometry of a specific cross section of a scanned object \cite{Baruchel2000,Desplentere2005}. For specimen I, a Tomographic image was obtained (figure \ref{fig:TEST-twill}) where it is possible to observe the high unbalance of the specimen and the characteristic weaving pattern of the 2.5 D interlock.

\begin{figure}[H]
\begin{centering}
\includegraphics[scale=1.3]{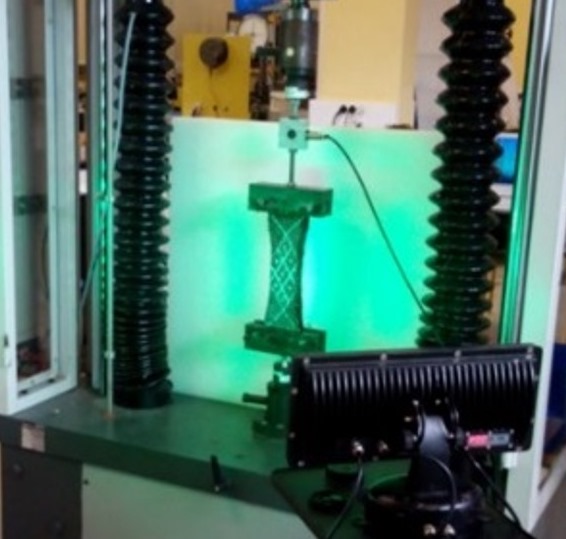}
\par\end{centering}

\protect\caption{Experimental device during testing \label{fig:TEST-device}}
\end{figure}

To conduct the BET, the sample is positioned between the upper and lower jaws of a 100 kN Zwick Meca tensile machine (figure \ref{fig:TEST-device}). The force needed to deform the specimen is on the order of 100 N and, therefore, an auxiliary load cell of 500 N is used to allow for good resolution of the data from the test. During the test, the lower clamp is static while the upper clamp is set, with a displacement-control, to move from 0 to 60 mm up. The displacement speed of the movable clamp is set to 4 mm/min. 

\begin{figure}[H]
\begin{centering}
\includegraphics[width=8cm,angle=0]{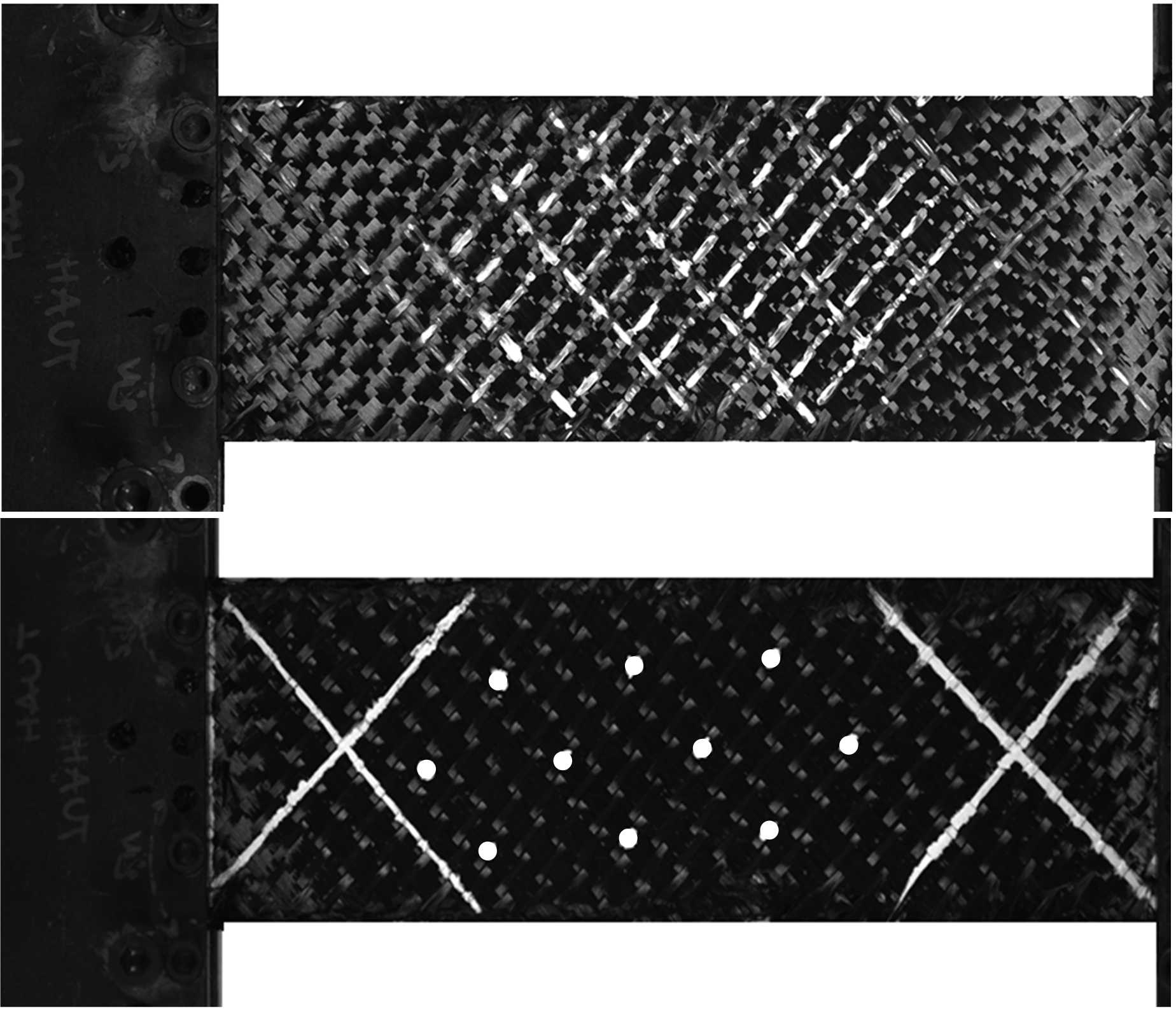}
\par\end{centering}
\protect\caption{(a) Sample I on top, (b) sample II at the bottom \label{fig:TEST-samples}}
\end{figure}

To analyze the experimental results and to reveal the characteristic deformation modes of the mesostructure, high-quality pictures of the samples during the deformation process are valuable. Therefore, a 16MP camera in combination with a 200x optical zoom and two LED adjustable color lights were used during the test. The detailed analysis of pictures of a black carbon specimen are still difficult to perform. For this reason, a white grid of lines aligned with the textile warp and weft yarns was added on the shear area A of the specimen I (figure \ref{fig:TEST-samples} (a)). However, during the analysis of the first specimen some difficulties following the deformation of the specimen were still present. Therefore, for specimen II, it was decided to locate only a couple of white points on top of area A that, with a post-processing step, lead to an easier access to important data like the local shear angle or the sliding between yarns. The initial configuration of the interlock specimens and the added reference lines and points are illustrated in figure \ref{fig:TEST-samples}.

\begin{figure}[H]
\begin{centering}
\includegraphics[scale=0.4]{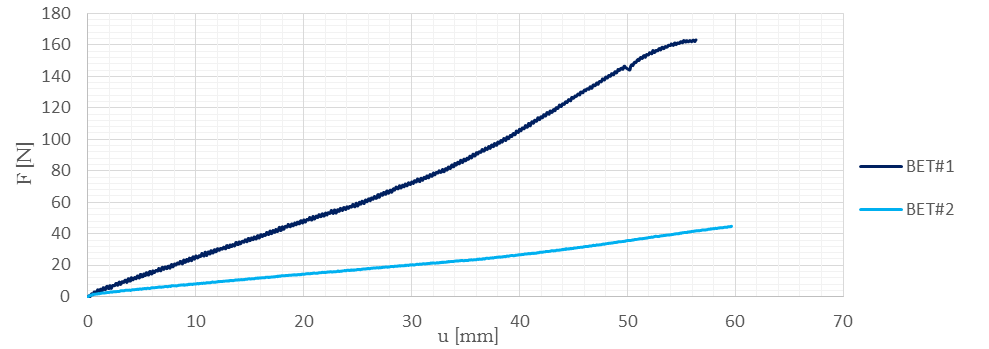}
\par\end{centering}

\protect\caption{Force/displacement plot\label{fig:TEST-FD}}
\end{figure}

The results of the tests in terms of the load-displacement curve are shown in figure \ref{fig:TEST-FD}. It can be seen that the force response for each of the specimens is almost linear except for a slight increase of stiffness at the end of the test. It is also easy to note that the two materials present very different macroscopic stiffnesses due to their different internal architecture.

\begin{figure}[H]
\begin{centering}
\includegraphics[width=11cm]{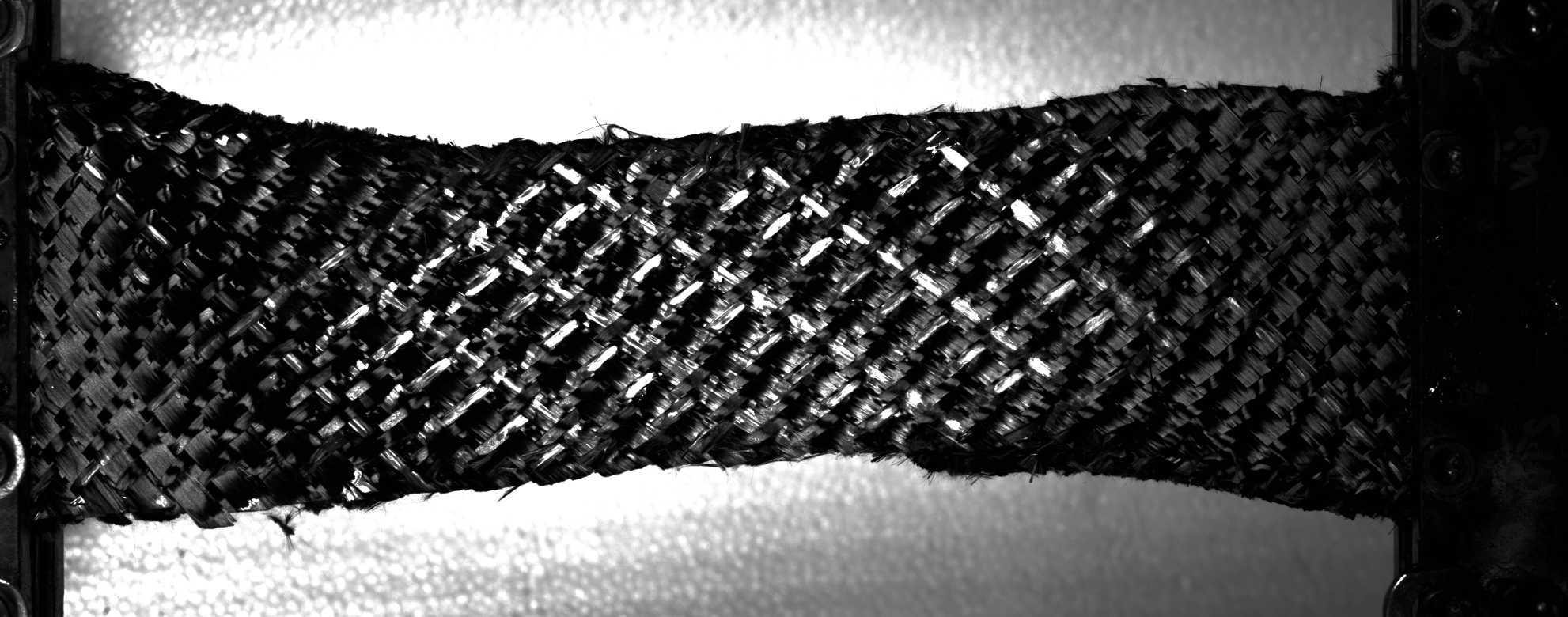}
\par\end{centering}

\begin{centering}
\includegraphics[width=11cm]{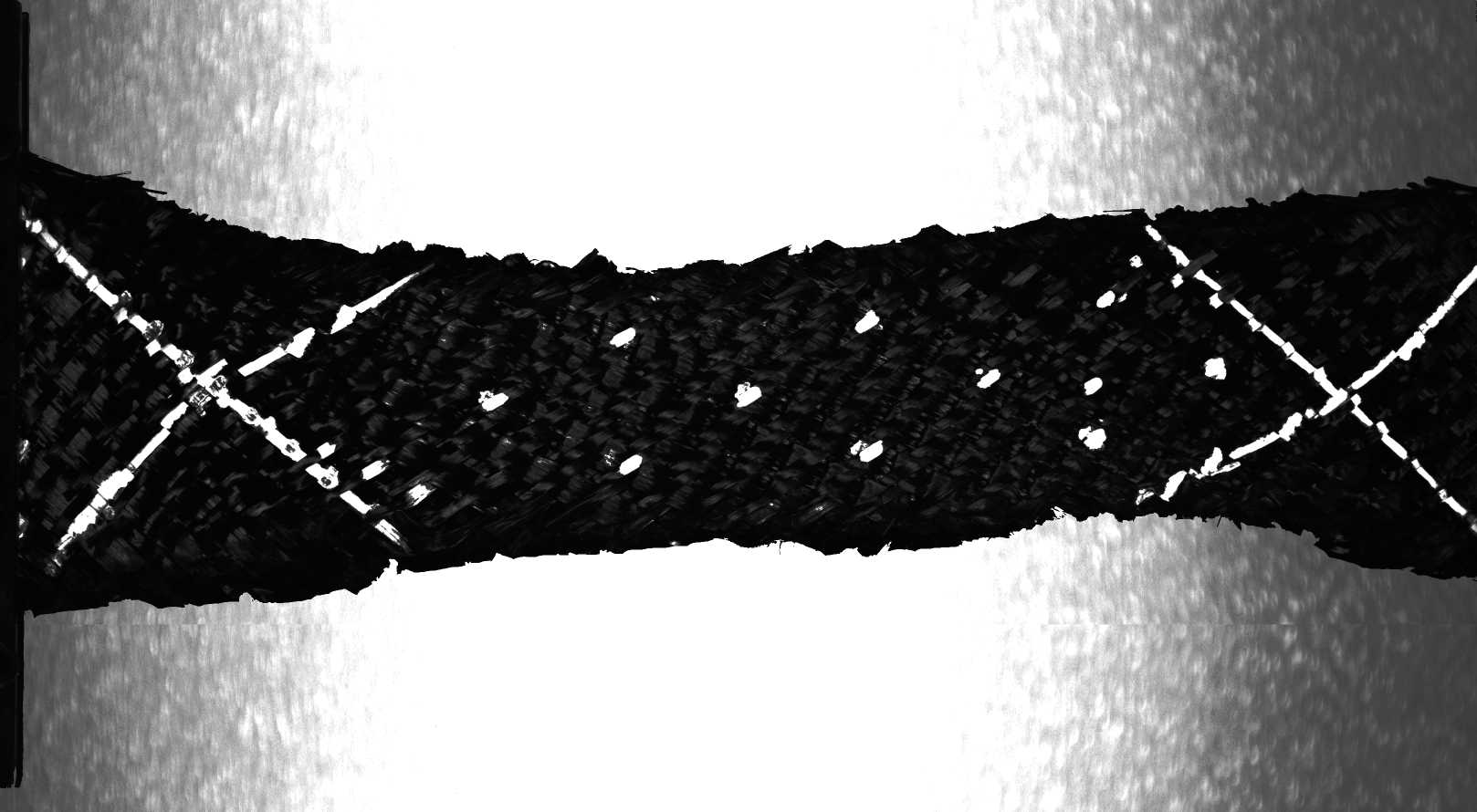}
\par\end{centering}

\protect\caption{Deformed shape for both specimens (sample I on top, sample II at the bottom) \label{fig:TEST-Shape}}
\end{figure}

Figure \ref{fig:TEST-Shape} shows the deformed shapes for both specimens during the development of the test. The quality of the test on the second specimen is much lower than that performed on the first one because some intrinsic asymmetries were introduced in the second specimen due to a non-perfect cutting. For this reason, the considerations will be illustrated by using the images relative to the specimen I, but analogous ones can be drawn for the specimen II. 

The main remark which may be inferred from the observation of the macroscopic deformed shape of the two specimens (figure \ref{fig:TEST-Shape}) is that it resulted in an asymmetric S-shape. This macroscopic asymmetry is not surprising considering the fact that the properties of the two families of yarns are very different in the two directions. What needs to be highlighted is the fact that such an asymmetric shape is related to precise deformation mechanisms of the meso-structure which have to be investigated to correctly understand the behavior of such unbalanced materials. In conclusion, a model, which represents with sufficient detail the the mechanical behavior of unbalanced fabrics, must be conceived in such a way to describe with sufficient accuracy:

\begin{itemize}
\item the macroscopic S-shaped deformation of the material,
\item the mesoscopic deformations of the yarns inside the material as related to the observed macroscopic S-shape.
\end{itemize}

To this goal, it is henceforth essential to observe what are the characteristic deformation patterns of the yarns inside the considered unbalanced fabric when subjected to a BET. As it will be better demonstrated in the remainder of this section, the main mesoscopic deformation mechanisms which take place during a BET performed on an unbalanced fabric are

\begin{itemize}
\item the in-plane shear deformation (angle variation between the yarns with respect to their initial configuration)
\item the local differential bending of the warp and weft yarns due to the unbalance of the fabrics
\item the relative slippage of the contact points between warp and weft yarns.
\end{itemize}

Ideally, if perfect pivots were placed to connect the warp and weft without interrupting the continuity of the yarns and if the two families of fibers could be modeled as wires with infinite rigidity with respect to elongation and vanishing bending stiffness, the observed motion would be the one presented in figure \ref{fig:BET-Schematics}. Thus, the only deformation mode would be the variation of the direction of the fibers which could be directly interpreted as the angle variation between warp and weft. Nevertheless, such an ideal situation is not the case in the considered material because the yarns present a non-vanishing bending stiffness and a relative slipping of the warp with respect to weft can also be observed.

\begin{figure}[H]
\begin{centering}
\includegraphics[width=15cm]{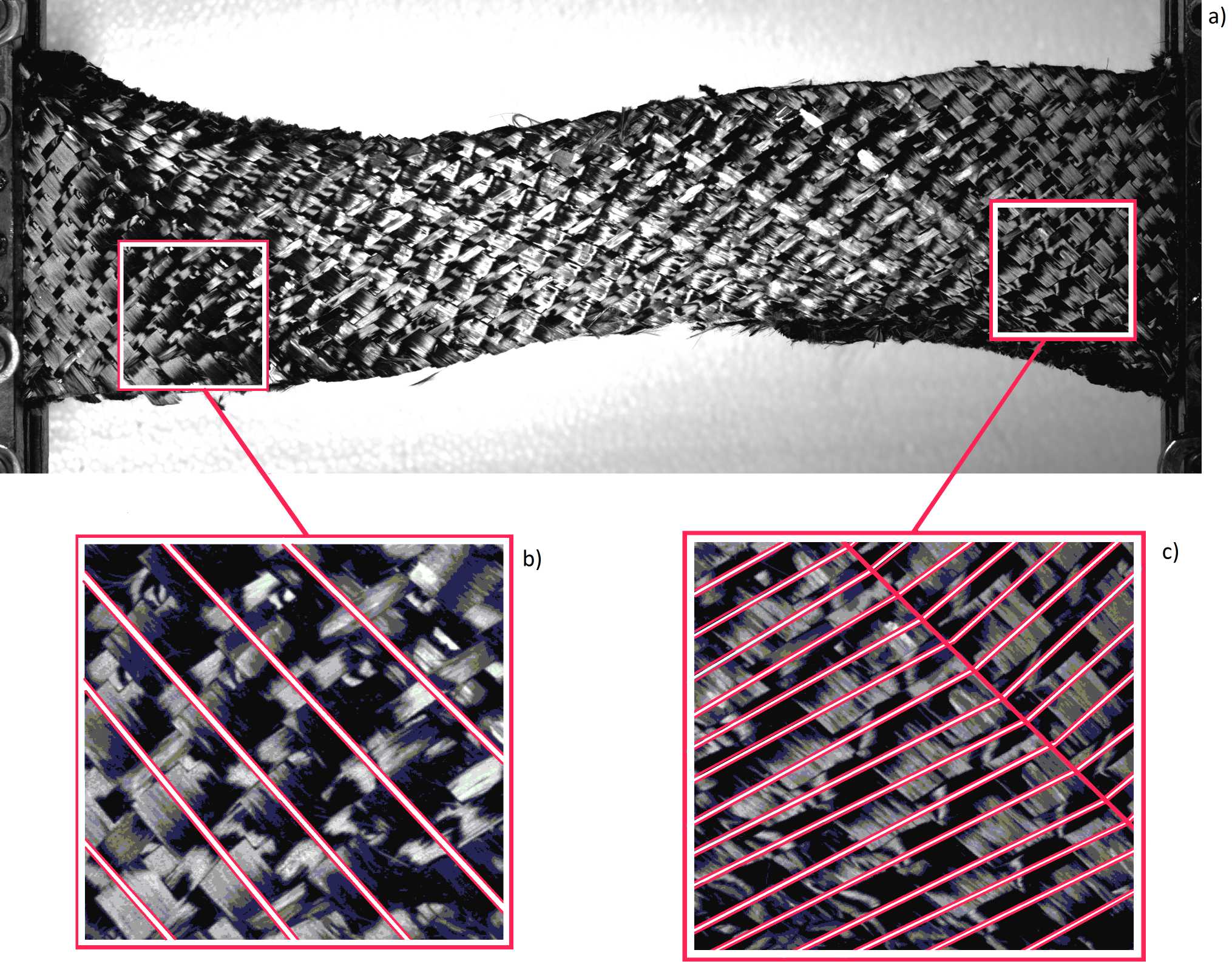}
\par\end{centering}

\protect\caption{Deformed shape for a displacement of 56 mm (a) and angle variation in the transition layers for the thick (b) and thin (c) yarns\label{fig:TEST-Bending}}
\end{figure}

More particularly, as far as the thin yarns are concerned, they possess a very low bending stiffness and, as it is possible to see in figure \ref{fig:TEST-Bending} (c), there is a very sharp variation of direction that can be considered to be concentrated in a very narrow layer. Instead, in the case of the thick yarns (figure \ref{fig:TEST-Bending} (b)), there is almost no measurable change in direction along the whole fiber, feature that can be uniquely related to an extremely high bending stiffness. If the two families of fibers are considered to be quasi-inextensible, the S-shape obtained in the test can be considered to be related to the very different bending stiffness of the two sets of yarns. As a consequence of this observation, it must be stated that it is not possible to describe this specific behavior without a model which accounts for the bending of the yarns at the mesoscopic level.

\begin{figure}[H]
\begin{centering}
\includegraphics[scale=0.2]{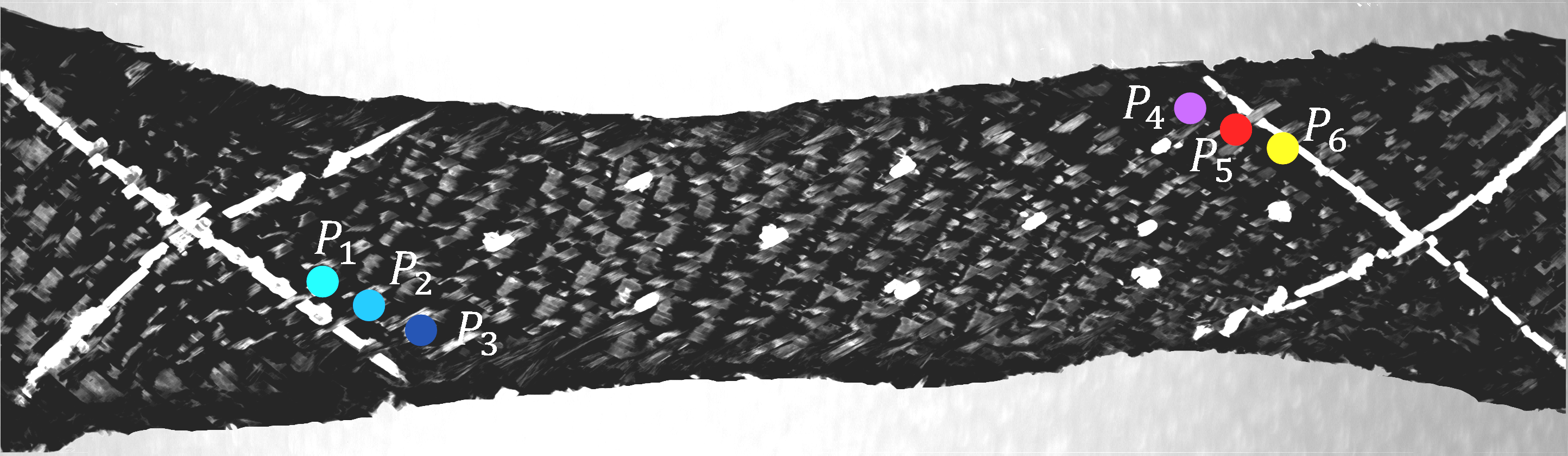}
\par\end{centering}

\protect\caption{Points identifying some sliding phenomena\label{fig:TEST-Slides}}
\end{figure}

It is finally noted that the presence of measurable slippages of the fibers (up to a maximum that is around 10\% of the total length of the yarns) strongly characterize this test. This phenomenon should be included in a complete model for this test and in general for woven composites. Such slippages can be qualitatively recognized when comparing figures \ref{fig:TEST-samples} (b) and \ref{fig:TEST-Shape} (b). It is possible to see that some points which were initially located on the white cross marks which were drawn on the specimen have moved, thereby breaking the continuity of the cross marks themselves (see also points $\mathrm{P}_{1}-\mathrm{P}_{6}$ in figure \ref{fig:TEST-Slides}).

A macroscopic indicator of the sliding may be found in the global in-plane thickness of the specimen. If relative sliding of the fibers is permitted, the height of the specimen measured in the middle of the specimen itself can be much higher than the height that the specimen would have if no relative motions were permitted. In some sense, such effect of the sliding can be modeled as a possible ``fictive elongation'' of the fibers in the two directions. More precisely, since the yarns can slide in the real situation, they can rearrange themselves in such a way that the resulting apparent in-plane thickness of the specimen is much higher than the one that the specimen would theoretically have if the yarns were hinged together. It is clear that the presence of such internal sliding weakens the basis on which a continuum theory is founded.

Nonetheless, it is possible to continue using a continuum model with the limit of modeling such internal sliding as a ``fictive'' elongation of the yarns in the two directions. The price to pay for this modeling assumption is that the result of the simulations is a microstructure which is not perfectly superposable to the real one (the ``real'' sliding is replaced by ``fictive'' elongations of the yarns). In spite of that price, the overall macroscopic pattern of the deformation is recovered, together with the main features of the deformation of the underlying microstructure. A model of this type could be of use for the simulation of the forming of unbalanced fabrics, even if the elongations which would be eventually present should be interpreted, at least partially, as relative sliding of the yarns. Finally, it must be explicitly remarked that the presence of the described relative sliding does not allow a direct interpretation of the in-plane shear deformation as the angle variation between the warp and weft directions. In fact, the in-plane shear is simply defined as the angle variation between the current direction of the considered yarn and its initial direction.

Summarizing, it is possible to say that the main mesoscopic deformation mechanisms which take place during the BET on an unbalanced fabric have been isolated. These significant contributions include the in-plane shear of the yarns, the differential local bending of the yarns and the sliding of the yarns. In the next section  a macroscopic, second-gradient, continuum model which is able to capture such mesoscopic deformation mechanisms together with their macroscopic counterpart  will be introduced. This model will provide a sensible description of the macroscopic S-shaped deformation of the considered unbalanced fabric, as well as a reasonable prediction of the mesoscopic deformation of the yarns.

\section{Second-Gradient Continuum Model}

As already remarked, to describe the response of the considered unbalanced fabric a second-gradient continuum model is introduced following the approaches of \cite{Phil1,Phil3D}. As long as no slipping occurs, two superimposed fibers can rotate around their contact point and, if a straight line is drawn on the textile reinforcement, one could see that it curves but remains continuous (see e.g. \cite{Boisse shaping}). As it has been established on a phenomenological basis, this situation is not always the case when performing a BET on unbalanced fabrics, because some relative sliding of the warp and weft yarns may occur during the test. The effect of such sliding is included in the overall behavior of the material by modeling it as a possible ``fictive elongation'' of the yarns, so reaching the threefold objective of 
\begin{itemize}
\item continue to use a continuum approach to model the mechanical behavior of unbalanced fabrics (it provides significant advantages in terms of engineering finite element modeling of the forming processes) 
\item obtain realistic macroscopic deformations for the considered unbalanced specimen
\item obtain realistic mesoscopic deformations of the yarns (the differential local bending of the fibers is precisely described and the sliding is accounted for by means of the introduction of fictive elongations).
\end{itemize}
Of course, the authors are aware of the fact that, such hypothesis of fictive elongation could be too restrictive for samples subjected to high strains, because it could produce some problems related to wrinkling during forming simulations. Nevertheless, given that only small-to-moderate strains are considered in this paper, the adopted continuum hypothesis can be  thought to represent a reasonable approximation of the underlying, microstructure-related, deformation mechanisms.

The main point in the continuous hyperelastic models is the definition of the strain energy density. Even if the energy is traditionally a function of only the strains (first-gradient theory) in different works such as \cite{Phil3D,dellIsSteig}, it is underlined that these types of energy may not be sufficient to model a class of complex contact interactions which are related to local bending stiffness of the yarns and which macroscopically affect the overall deformation of interlocks. As it has been previously highlighted, the phenomenon of the local bending of the yarns is a mesoscopic deformation pattern when applying the BET to an unbalanced fabric. For this reason, a model which is able to correctly account for the different bending stiffnesses of the two families of yarns must include second-gradient effects. As a matter of fact, a first-gradient continuum model in which very different tension stiffnesses of the yarns in the two directions are considered would allow the reproduction of an asymmetric macroscopic S-shape of the specimen. Nevertheless, the description of the deformation of the single yarns would be unrealistic: sensible differential elongations of the two sets of yarns and piecewise straight lines would be the only possible deformed shapes for the yarns constituting the fabric. Even if an ``ad hoc'' highly anisotropic choice of the constitutive parameters would allow reproduction of the desired macroscopic asymmetric S-shape, the material parameters would not be  representative of the actual material behavior.  This phenomenon will be explicitly shown in section \ref{subFirstGradient}.

In what follows, a second-gradient energy model that takes into account, in an averaged sense, some micro-structural properties such as the bending of the fibers will be presented with the same spirit of what done e.g. in \cite{NthGrad,BC2grad1,BC2grad2,BC2grad3,Germain}.
The second-gradient model proposed here is intrinsically macroscopic, in the sense that it is not easy to directly relate the introduced macroscopic coefficients to precise microscopic characteristics of the fibers. To this aim, suitable bottom-up approaches for networks of fibers should be used following the ideas presented e.g. in \cite{Alibert,Seppecher Exotic,dellIsSteig,David7,David1,David2,David3}.

\subsection{Hyperelastic Orthotropic Second Gradient Strain Energy Density}

When working with a continuum model, it is important to introduce a Lagrangian configuration $\ensuremath{B_{L}\subset\mathbb{R}^{3}}$ and a suitably regular kinematic field $\ensuremath{\boldsymbol{\chi}(\mathbf{X},t)}$ which associates to any material point $\ensuremath{\mathbf{X}\in B_{L}}$ its current position $\ensuremath{\mathbf{x}}$ at time t. The image of the function $\ensuremath{\ensuremath{\boldsymbol{\chi}}}$ gives, at any instant t, the current shape of the body $\ensuremath{B_{E}(t)}$: this time-varying domain is usually referred to as the Eulerian configuration of the medium and, indeed, it represents the system during its deformation. Because they will be used in the following, the displacement field $\ensuremath{\mathbf{u}(\mathbf{X},t):=\boldsymbol{\chi}(\mathbf{X},t)-\mathbf{X}}$, the tensor $\ensuremath{\mathbf{F}:=\nabla\boldsymbol{\chi}}$ and the Right Cauchy-Green deformation tensor\footnote{A central dot indicates simple contraction between tensors of order greater than zero. For example if $\ensuremath{\mathbf{A}}$ and $\ensuremath{\mathbf{B}}$ are second order tensors of components $\ensuremath{A_{ij}}$ and $\ensuremath{B_{jh}}$ respectively, then $\ensuremath{\ensuremath{(\mathbf{A}\cdot\mathbf{B})_{ih}:=A_{ij}B_{jh}}}$, where Einstein notation of sum over repeated indexes is used.} $\ensuremath{\mathbf{C}:=\mathbf{F}^{T}\cdot\mathbf{F}}$ are introduced. 
The first-gradient kinematics of the continuum must be enriched by considering the second order tensor field $\nabla\ensuremath{\mathbf{\mathbf{C}}}$ which accounts for terms that can be associated to the macro-inhomogeneity of micro-deformation in the microstructure of the continuum. This model could be considered as a limiting case of generalized continua with microstructure like the one presented in \cite{Mindlin} for the linear-elastic case and those of \cite{EringenI,Eringen II,Forest0,Forest} for the case of non-linear elasticity. Because a second-gradient theory can be readily obtained as limiting case of the micromorphic theory, one can then derive the second-gradient contact actions in terms of the micromorphic ones following the procedure used in \cite{Bleustein}. Some of the possible types of constraints that could be included in the proposed micromorphic model which, for example, impose inextensibility of yarns giving rise to so-called micropolar continua are presented in \cite{Victor3,Victor6,Victor00,Victor5,EringenBook}. 

A hyperelastic, orthotropic, second-gradient model can be applied to the case of relatively thin fibrous composite reinforcements at finite strains. For the strain energy density $\Psi\ensuremath{\left(\mathbf{C},\ \nabla\mathbf{C}\right)}$ which shall be used to simulate the mechanical behavior of the fibrous composite reinforcements in the finite strain regime it is assumed a decomposition such as:

\begin{equation}
\Psi\left(\mathbf{C},\ \nabla\mathbf{C}\right)=\Psi_{I}(\mathbf{C})+\Psi_{II}(\nabla\mathbf{C}). \label{E}
\end{equation}

In this formula, $\ensuremath{\Psi_{I}}$ is the first-gradient strain energy and $\ensuremath{\Psi_{II}}$ is the second-gradient strain energy. In this work, only an elastic analysis will be considered neglecting the phenomena such as the damage that can be analyzed in a second-gradient framework as done in \cite{Placidi1,Placidi2}. The additive decomposition of the strain energy density of Eq. \eqref{E} is based on the assumption that first and second-gradient effects are uncoupled in the considered problem.

Explicit expressions for isotropic strain energies suitable for modeling isotropic materials even at finite strains are available in literature (see e.g. \cite{Ogden,David6}). For linear elastic isotropic second-gradient media, it is possible to determine generalized constitutive laws (see \cite{Hooke2grad}). Instead, in the case of orthotropy, strain energy potential expressions suitable to describe the real behavior of such materials are not common in the literature. For orthotropic materials, certain constitutive models are for instance presented in \cite{Itskov}, where some polyconvex energies are proposed to describe the deformation of rubbers in uniaxial tests. Explicit anisotropic hyperelastic potentials for soft biological tissues are also proposed in \cite{Holzapfel} and reconsidered in \cite{Neff1,NeffBis} in which their polyconvex approximations are derived. Other examples of polyconvex energies for anisotropic solids are given in \cite{David5}. 

It is even less common to find in the literature reliable constitutive models for the description of the real behavior of fibrous composite reinforcements at finite strains, but some attempts can be for instance recovered in \cite{Aimene,Phil1}. Furthermore, the mechanical behavior of composite preforms with a rigid organic matrix (see e.g. \cite{Yves1,Yves3,Yves4,Yves2}) is quite different from the behavior of the sole fibrous reinforcements (see e.g. \cite{Phil3D}) rendering the mechanical characterization of such materials a major scientific and technological issue. 

In this work, the preferred directions $\ensuremath{\mathbf{m}_{1}}$ and $\ensuremath{\mathbf{m}_{2}}$ are considered as the initial directions of the yarns and the direction $\ensuremath{\mathbf{m}_{3}}$ as the normal to the plane of the woven fabric in the reference configuration. For relatively thin interlocks loaded in the plane, the out-of-plane effects can be considered to be negligible, therefore it is possible to use an orthotropic energy with an expression of the type:

\begin{equation}
\Psi_{I}(\mathbf{C})=\Psi_{I}(i_{4},i_{6},i_{8}),
\end{equation}

where the in-plane invariants of the Cauchy-Green deformation tensor $\mathbf{C}$ are defined as:

\begin{itemize}
\item $i_{4}\text{=}\mathbf{m}_{1}\cdot\mathbf{C}\cdot\mathbf{m}_{1}$ represents
the elongation strain in the direction $\ensuremath{\mathbf{m}_{1}}$ 
\item $i_{6}\text{=}\mathbf{m}_{2}\cdot\mathbf{C}\cdot\mathbf{m}_{2}$ represents
the elongation strain in the direction $\ensuremath{\mathbf{m}_{2}}$ 
\item $i_{8}\text{=}\mathbf{m}_{1}\cdot\mathbf{C}\cdot\mathbf{m}_{2}$ represents
the shear strain between the directions $\ensuremath{\mathbf{m}_{1}}$
and $\ensuremath{\mathbf{m}_{2}}$ that can be related to the angle
variation
\end{itemize}

These invariants of the tensor $\mathbf{C}$ can be used to describe the main first-gradient deformation mechanisms intervening during the considered test that are the in-plane angle shear deformation of the yarns and the slippages of the two families of fibers. The energy associated with the shear angle variation can be described as a function of $i_{8}$ while the elongation parameters $i_{4}$ and $i_{6}$ are used to model the presence of the slippage by means of an equivalent continuum model. It is possible to develop complex non-linear energies, but that development is not one of the aims of this paper. Instead, a simple first-gradient energy is introduced to describe the overall behavior of the considered material, such as:

\begin{equation}
\Psi_{I}(\mathbf{C})=\frac{1}{2}K_{el}\left[(\sqrt{i_{4}}-1)^{2}+(\sqrt{i_{6}}-1)^{2}\right]+\frac{1}{2}K_{sh}i_{8}^{2}
\end{equation}

This energy introduces an extensional equivalent stiffness $K_{el}$ and a shear stiffness $K_{sh}$. The first two terms account for the equivalent elongations of the yarns in the warp and weft directions, respectively. As already mentioned, such an elongation is fictitiously used to model the slipping of the yarns one with respect to the other. Such approximation can be considered to be sensible because the overall effect of the slipping at the mesoscopic level results in a macroscopic increase of the width of the deformed specimen. Such a width increase is fictitiously interpreted as an ``elongation'' of the yarns in the considered continuum model. The last term accounts for the shear deformation: angle variation between the yarns in the current configuration with respect to the reference one. It must be repeated once again that the extensional effects included in the model are associated with the description of the slipping of the two families of fibers while the shear term models the energy associated with the in-plane shear deformation. Because the two extensional stiffness parameters describe the mutual interaction between the two families of yarns, and it is reasonable that the friction properties remain the same for the two directions, the simplification of the same extensional stiffness in the two directions of the fabrics was assumed. Furthermore, with such a simplification, the formulation has a lower number of parameters and the asymmetry of the material can be completely attributed to the different values of the second-gradient parameters in the two directions. 

The main observable experimental phenomenon that cannot be included by means of a classical first-gradient theory is the bending stiffness of the yarns. On the other hand, such local bending of the yarns can be included by introducing a second-gradient energy function of the space derivatives of the invariant $i_{8}$, rough descriptors of the in-plane curvature of the yarns of the fabric. The constitutive choice of the second-gradient energy for the considered unbalanced fabrics is of the type

\begin{equation}
\Psi_{II}(\mathbf{\nabla\mathbf{C}})=\Psi_{II}(\nabla i_{8})=\Psi_{II}(i_{8,1},i_{8,2}),
\end{equation}

where $i_{8,1}$ and $i_{8,2}$ are the derivatives of $i_{8}$ with respect to the space coordinates along $\ensuremath{\mathbf{m}_{1}}$
and $\ensuremath{\mathbf{m}_{2}}$. Once again, a simple form was chosen even for the second-gradient energy and, furthermore, because the fabric is strongly unbalanced, the coefficient of the two derivatives were considered distinct leading to an energy of the type:

\begin{equation}
\Psi_{II}(\nabla\mathbf{C})=K_{3} \, i_{8,1}^{2}+K_{4} \, i_{8,2}^{2}.\label{eq:2GEn}
\end{equation}

This energy introduces the local bending stiffness $K_{3}$ of the yarns in direction $\ensuremath{\mathbf{m}_{1}}$ and the local bending stiffness $K_{4}$ of the yarns in direction $\ensuremath{\mathbf{m}_{2}}$ into the model. The second-gradient hyperelastic model proposed in this paper is based on a phenomenological approach: the addition of the second-gradient terms in the strain energy density as specified in Eq. \eqref{eq:2GEn} allows us to describe with a reasonable accuracy the onset of the observed phenomena. Summarizing, the final form of the energy considered here takes the form:

\begin{equation}
\Psi\left(\mathbf{C},\nabla\mathbf{C}\right)=\frac{1}{2}K_{1}\left[(\sqrt{i_{4}}-1)^{2}+(\sqrt{i_{6}}-1)^{2}\right]+\frac{1}{2}K_{2}i_{8}^{2}+\frac{1}{2} \, \nabla i_8\cdot \mathbf K \cdot \nabla i_8 \label{eq:SecGradConstitutive}
\end{equation}

where it was set
$$
\mathbf K =\left( \begin{array}{cc}
K_3 & 0  \\ 
 0 & K_4 
\end{array}
\right).
$$

\subsection{Governing equations in strong form}\label{StrongForm}
The governing equations in strong form and the associated  boundary conditions of the considered generalized continuum can be a useful tool for the understanding of the physics of the considered problem. In particular, the boundary conditions which can be assigned in the considered problem describe the possible interactions of the external world with the considered specimen as it will be shown in the remainder of this section. Following what done in \cite{MadeoBias}, it is convenient to pass through a reformulation of the problem which makes uses of a constrained micromorphic model. More precisely, instead of minimizing a problem in which a second-gradient energy of the type  (\ref{eq:SecGradConstitutive}) is considered, a supplementary kinematical field $\psi$ and a Lagrange multiplier $\Lambda$ are introduced such that the strain energy density can be written as

\begin{equation}
\widetilde \Psi\left(\mathbf{C},\nabla \psi, \Lambda \right)=\frac{1}{2}K_{1}\left[(\sqrt{i_{4}}-1)^{2}+(\sqrt{i_{6}}-1)^{2}\right]+\frac{1}{2}K_{2}i_{8}^{2}+\frac{1}{2} \nabla \psi \cdot \mathbf K \cdot \nabla \psi+ \Lambda \ (\psi-i_8).\label{eq:SecGradConstitutive_microm}
\end{equation}
A minimization problem of such a micromorphic model leads to the following set of bulk equations
$$
\mathrm{Div} \, \boldsymbol \sigma = 0, \qquad \mathrm{Div} \ \mathbf M + \Lambda=0, \qquad \psi-i_8=0,
$$
with 
$$
\boldsymbol \sigma = \mathbf F \cdot \left[ K_1  \frac{\sqrt{i_4}-1}{\sqrt{i_4}} (\mathbf m_1 \otimes \mathbf m_1) + K_1  \frac{\sqrt{i_6}-1}{\sqrt{i_6}} (\mathbf m_2 \otimes \mathbf m_2) + (K_2 \, i_8-\Lambda) \,  (\mathbf m_1 \otimes \mathbf m_2+ \mathbf m_2 \otimes \mathbf m_1)\right]
$$
and
$$
\mathbf M = \mathbf K \cdot \nabla \psi.
$$
On the other hand, the boundary conditions that can be assigned on the boundaries of the considered specimen are 
\begin{equation} \boldsymbol \sigma \cdot \mathbf n = \mathbf f^{ext}, \qquad \text{OR} \qquad  \mathbf u = \mathbf u^{ext} \label{BC1}
\end{equation}
and
\begin{equation} \mathbf M \cdot \mathbf n = g^{ext}, \qquad \text{OR}  \qquad \psi = \psi^{ext}.  \label{BC2}
\end{equation}

If now one lets $\psi$ tend to the angle variation $i_8$, the second-gradient model presented in the previous subsection can be recovered. The implementation of a second-gradient theory through a constrained micromorphic theory in which $\psi$ is set to be equal to $i_8$ allows a better understanding of the boundary conditions which take an intuitive meaning. In the proposed model, and with reference to Eqs. \eqref{BC1} and \eqref{BC2}, can be assigned on the boundaries of the considered specimen the force $\mathbf f^{ext}$ or the displacement $\mathbf u^{ext}$ AND the couple $g^{ext}$ or the angle variation $\psi^{ext}$.

The boundary conditions that are used in the numerical simulations proposed in the next subsection are:
\begin{itemize}
\item vanishing displacement on one clamped edge and imposed displacement on the second clamped edge,
\item vanishing angle variation at the two clamped hands (the angle between the yarns does not vary due to the clamp),
\item vanishing forces and couples on the two free boundaries.
\end{itemize}

\subsection{Numerical Results\label{sec:Numerical-Model}}

\begin{figure}[H]
\begin{centering}
\includegraphics[scale=0.18]{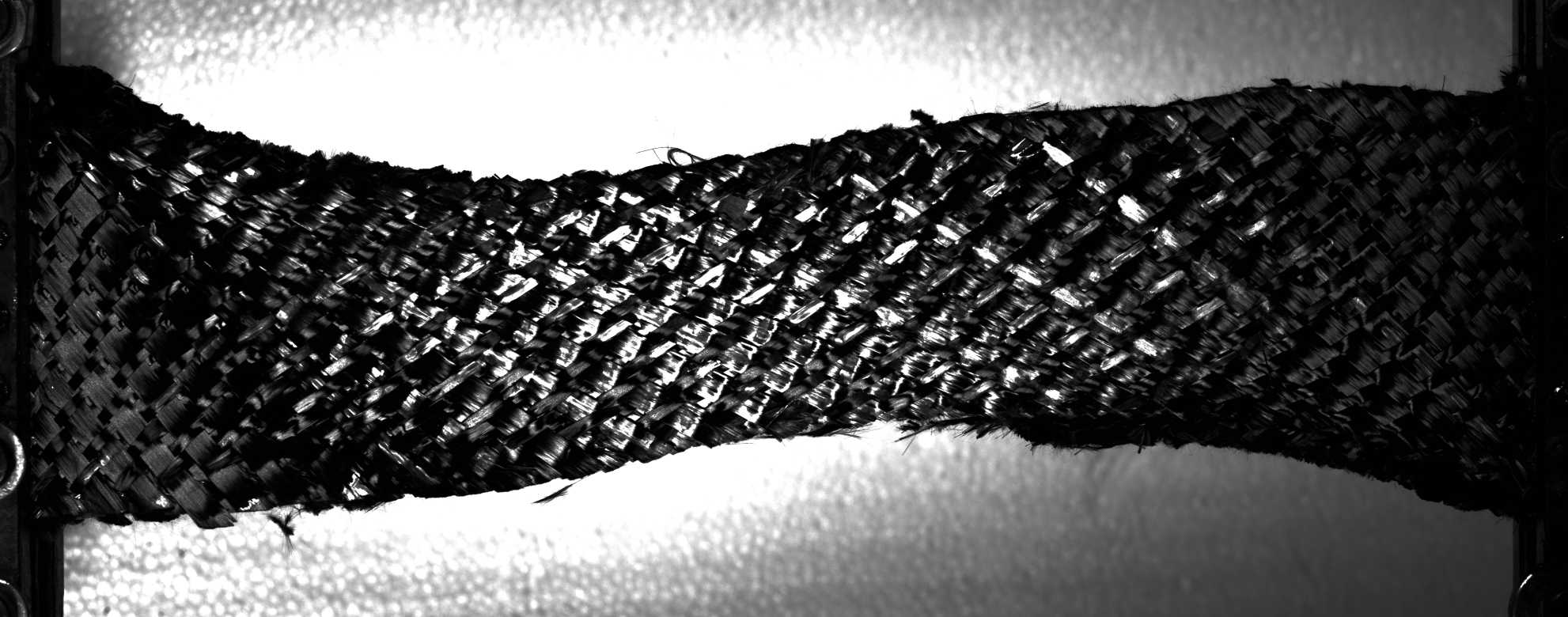}
\par\end{centering}

\protect\caption{Experimental Shape for a Displacement of 56 mm \label{fig:TEST-56}}
\end{figure}

In this section, the results obtained with the introduced hyperelastic, unbalanced second-gradient model are shown. To clearly evaluate the obtained results, it is useful to show once again the experimental shape (figure \ref{fig:TEST-56}) as reference for all of the following considerations. 

 It is reminded to the reader, that the aim is not to fit the proposed model on the material II due to some problems that occurred during the cutting of the specimen and, thus, could have affected the behavior of the material in a non-perfectly controllable manner.

\begin{figure}[H]
\begin{centering}
\includegraphics[scale=0.5]{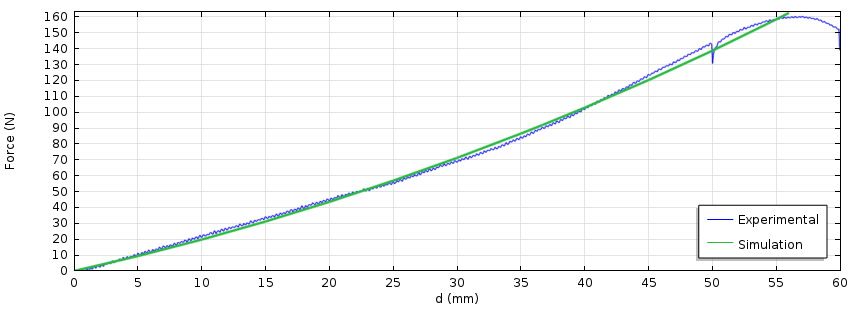}
\par\end{centering}

\protect\caption{Experimental and numerical load-displacement curves for the Specimen
I of the considered unbalanced interlock. \label{fig:SIM-LD}}
\end{figure}

The constitutive parameters appearing in Eq. \eqref{eq:SecGradConstitutive} were heuristically chosen by using an inverse method based on physical observations. More particularly, the values of the second-gradient parameters were initially chosen in such a way that the bending stiffness of the thin yarns $K_{4}$  is very small (eventually almost vanishing), while the bending stiffness  of the thick yarns $K_{3}$ is subsequently chosen to fit at best the experimental S-shape of the specimen. Notwithstanding the values given to the parameters $K_{3}$ and $K_{4}$ the complete macroscopic S-shape of the specimen cannot be recovered without tuning the value of the ``sliding'' parameter $K_{1}$ which is seen to have a direct influence on the in-plane thickness of the specimen. Activating this parameter, the height of the specimen in the middle of the specimen itself starts increasing and becomes closer to the experimental shape. A subsequent parametric study  is made on the values of the quoted parameters so as to reach the best possible fitting of the experimental S-Shape. A different treatment is left to the in-plane shear parameter $K_{2}$ whose value is tuned to fit the experimental load-displacement curve (figure \ref{fig:SIM-LD}).

A suitable optimization technique could be set up to recover the values of the parameters which better fit the experimental evidence. Although interesting, this point is left open for further works which will be based on a more representative amount of experimental data. Here, it is only presented a more intuitive calibration of the parameters which nevertheless allows to unveil the single effect of each of them on the macroscopic deformation of the considered specimen.

The second-gradient hyperelastic model was implemented in $\text{COMSOL}^{\text{\ensuremath{\circledR}}}$ to model the behavior of the first specimen, and the parameters used in the simulations are shown in the Tab. \ref{table: Par}. 
Such implementation has been performed using the ``weak form'' package in which the expression for the power of internal actions associated to the energy in Eq. \eqref{eq:SecGradConstitutive_microm} has been explicitly given by imposing the constraint $\psi=i_8$ via a suitable Lagrange multiplier. Moreover, the boundary conditions used are those explicitly mentioned at the end of section \ref{StrongForm}.

\begin{table}[H]
	\begin{centering}
		\begin{tabular}{|c|c|c|c|}
			\hline 
			$K_{1}$ & $K_{2}$ & $K_{3}$ & $K_{4}$\tabularnewline
			\hline 
			\hline 
			0.7 MPa & 21 kPa & 2 kPa$\cdot$m$^{2}$ & 0.02 kPa$\cdot$m$^{2}$\tabularnewline
			\hline 
		\end{tabular}
		\par\end{centering}
	
	\protect\caption{Parameters of the proposed second-gradient continuum model\label{table: Par}}
\end{table}

\begin{figure}[H]
\begin{centering}
\includegraphics[scale=0.4]{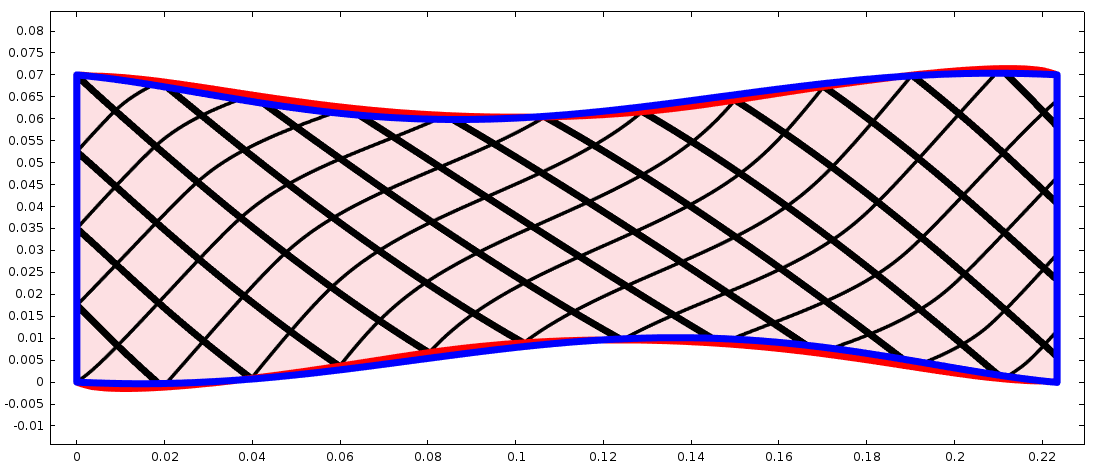}
\par\end{centering}

\protect\caption{Deformed shape in the simulation (red with black fibers) and experimental (blue) for an imposed displacement of 37 mm \label{fig:SIM-Shape}}
\end{figure}

\begin{figure}[H]
\begin{centering}
\includegraphics[scale=0.4]{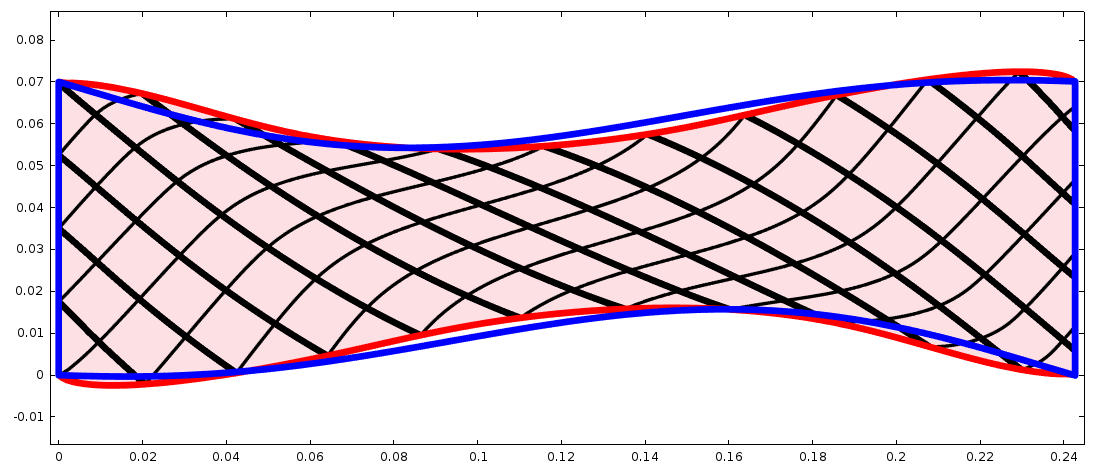}
\par\end{centering}

\protect\caption{Deformed shape in the simulation (red with black fibers) and experimental (blue) for an imposed displacement of 56 mm \label{fig:SIM-Shape-1}}
\end{figure}

Figures \ref{fig:SIM-Shape} and \ref{fig:SIM-Shape-1} show the results of the proposed second-gradient model as compared with the experimental shape (in blue) and for two different values of the displacement imposed at one edge. The two sets of fibers are drawn inside the specimen with different thicknesses to schematically represent the unbalance of the fabric. It can be inferred that the second-gradient simulations match the experimental response very well even for different values of the imposed displacement. Moreover, internal deformation of the mesostructure can be considered to be reasonable because the fiber lines are coherent with the experimental evidence and even the width of the specimen matches the experimental results. 

The reader is cautioned that the fiber lines obtained with the simulation, and shown in the figures \ref{fig:SIM-Shape} and \ref{fig:SIM-Shape-1}, do not precisely represent the real patterns of the microstructure. The slipping phenomenon is included in the model through the introduction of ``fictive'' elongations which of course allow an averaged macroscopic description of the slipping itself but not a precise determination of microscopic relative displacements of warp and weft.

\begin{figure}[H]
\begin{centering}
\includegraphics[scale=0.5]{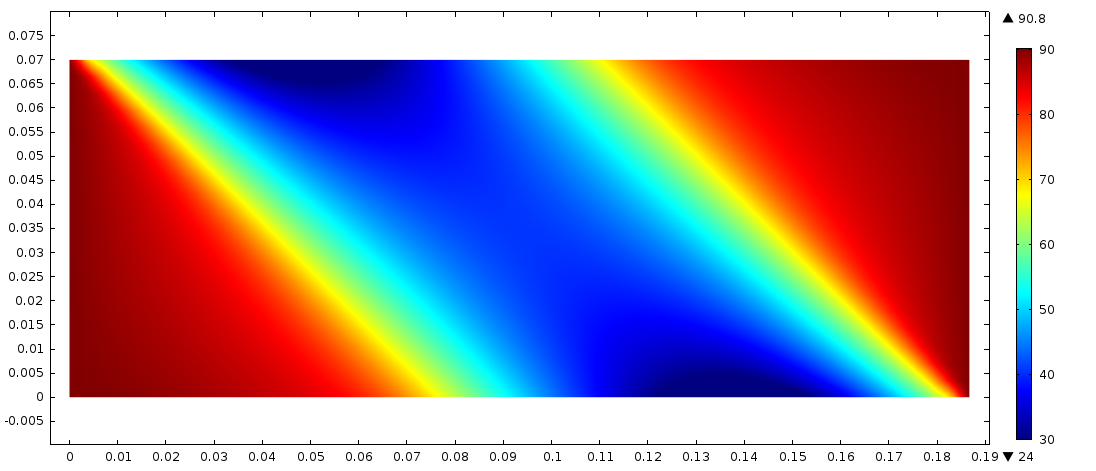}
\par\end{centering}

\protect\caption{Angles between the Fibers in the Second Gradient Simulation for a Displacement of of 56 mm \label{fig:SIM-AngleVariation}}
\end{figure}

As it is usually done when exploiting the results of a Bias Extension Test, the pattern of the angle between the yarns is also investigated. It is possible to see in figure \ref{fig:SIM-AngleVariation} that the angle between the yarns is almost constant in three regions which are separated by thin transition layers aligned along the strong yarns' direction. This result is in agreement with the available experimental evidence and is due to the high bending stiffness of the thick yarns that find energetically convenient the fact of remaining almost straight instead of varying their direction as happens in a standard, balanced Bias Extension Test. 

Finally, it is interesting to evaluate the overall response of the model in terms of forces. As previously remarked, suitably tuning the shear stiffness $K_{2}$ until reaching the value shown in Tab. \ref{table: Par} allows us to obtain the force-displacement curve shown in figure \ref{fig:SIM-LD}. The slight non-linearity in the simulated response, that partially reflects the experimental increase of stiffness, is due only to the geometric non-linearity of the problem while the constitutive laws of the problem are linear with respect to the introduced invariants. Furthermore, the approximation of the slippage as an elongation does not affect the ability to describe the global response of the specimen as well as the main features of the response at the mesoscopic level. 

It is explicitly mentioned the fact that in the presented simulations the force has been calculated according to Eq. \eqref{BC1}.

\subsection{First-gradient models are insufficient to model the BET on unbalanced fabrics} \label{subFirstGradient}

In this concluding subsection, the fact that first-gradient models are not able to capture the complex behaviors of the microstructures is highlighted. As a matter of fact, one could, at a first glance, believe that the asymmetric S-shape of the macroscopic specimen, which can be observed during the BET on unbalanced fabrics, could actually be reproduced by using a simple first-gradient theory in which very different material properties are used in the two directions. This approach would mean that it is possible to reproduce the observed phenomena by choosing an energy of the type
\begin{equation}
\Psi_{I}\left(\mathbf{C},\nabla\mathbf{C}\right)=\frac{1}{2}A_{1}(\sqrt{i_{4}}-1)^{2}+\frac{1}{2}B_{1}(\sqrt{i_{6}}-1)^{2}+\frac{1}{2}K_{2}i_{8}^{2},\label{eq:FirstGradConstitutive}
\end{equation}
with very different values of the constants $A_{1}$ and $B_{1}$ from the coefficients used in Eq. \eqref{eq:SecGradConstitutive}.

\begin{figure}[H]
\begin{centering}
\includegraphics[scale=0.4]{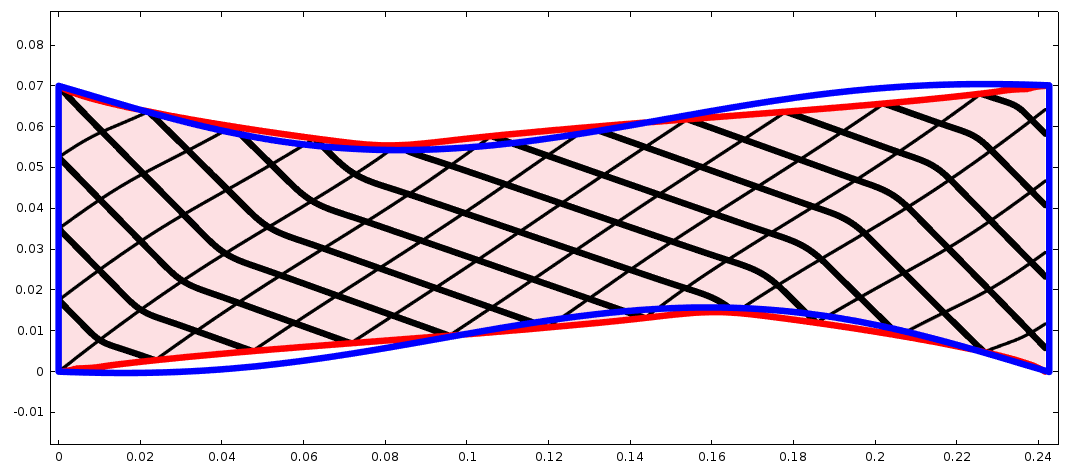}
\par\end{centering}

\protect\caption{\label{fig:First-gradient-solution}First-gradient solution obtained with the constitutive choice of the parameters given is Tab. \ref{table: Par-2}.}
\end{figure}

Indeed, when suitably tuning the coefficients appearing in Eq. \eqref{eq:FirstGradConstitutive} by choosing $A_{1}$ and $B_{1}$ with significantly different values, it is true that an asymmetry can be produced in the macroscopic shape of the specimen which agrees with the macroscopic experimental S-shape. Nevertheless, such macroscopic shape is not associated to any reasonable motions of the yarns at the mesoscopic level. More precisely, if the first-gradient coefficients appearing in the energy (Eq. \eqref{eq:FirstGradConstitutive}) are chosen to take the values given in Tab. \ref{table: Par-2}, then the solution shown in figure \ref{fig:First-gradient-solution}  is obtained.

\begin{table}[h]
\begin{centering}
\begin{tabular}{|c|c|c|}
\hline 
$A_{1}$ & $B_{1}$ & $K_{2}$\tabularnewline
\hline 
\hline 
0.55 MPa & 110 MPa & 0.11 MPa\tabularnewline
\hline 
\end{tabular}
\par\end{centering}

\protect\caption{Parameters of the first-gradient continuous model \label{table: Par-2}}
\end{table}

 It is possible to notice from the first-gradient results shown in figure \ref{fig:First-gradient-solution} that the obtained deformation pattern turns to be completely unphysical for the following reasons:
\begin{itemize}
\item even if an asymmetry of the macroscopic shape can be obtained, the boundary of the specimen obtained by means of the first-gradient model is piecewise linear, so that the actual curvature of the specimen cannot be precisely recovered; 
\item the pattern of the microstructure associated with the desired macroscopic shape is completely unrealistic: it is possible to see that the thick yarns bend while the thin ones stay straight. Moreover the variation of direction associated with such unphysical bending is concentrated on a very thin region, which is sensible in a first-gradient theory where no bending stiffness is associated with the yarns.
\end{itemize}

\begin{figure}[H]
\begin{centering}
\includegraphics[scale=0.4]{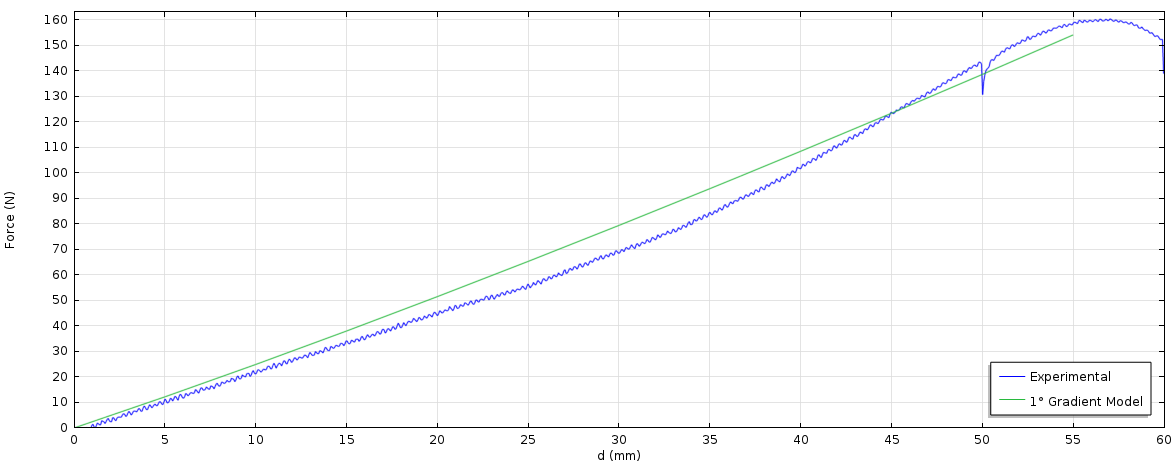}
\par\end{centering}
\protect\caption{\label{fig:First-gradient-load} Force-displacement curve obtained with the first-gradient model.}
\end{figure}

Moreover, it is shown in figure \ref{fig:First-gradient-load} that the limits of the considered first-gradient model can be unveiled also with reference to the load-displacement curve which significantly differs from the experimental one.

The chosen first-gradient constitutive expression could be replaced with any other first-gradient one (for example Mooney-Rivlin or Ogden), but the internal trend of the fibers could never be recovered due to the fact that, by definition, no second order derivatives of displacement are present in first-gradient models. Indeed, such higher order derivatives can be directly related to the curvature of the fibers and it is thus possible to conclude that only a second-gradient constitutive law allows to explicitly account for the (differential) bending of the yarns and, hence, for the description of the real deformation patterns of the fibers.

The results presented in this last subsection show with no doubt that second-gradient theories are necessary to correctly model the differential bending stiffness of the two sets of yarns in a physically sound way while remaining in a continuum framework.

\section*{Conclusions}

The traditional continuum models based on the implementation of first-gradient energies neglect the description of some essential microstructure-related physical phenomena in woven fibrous composite reinforcements, such as the local bending of the yarns. The insertion of higher order terms in the energy is unavoidable if one wants to correctly describe such kind of materials in the framework of macroscopic continuum theories. In particular, in this paper the results obtained with the use of a suitable second-gradient energy that describe some particular experimental behaviors of woven fabrics were presented, namely: 
\begin{itemize}
\item the in-plane shear deformation
\item the asymmetry of the macroscopic shape due to the unbalanced bending stiffness of the warp and weft yarns
\item the slippage of warp yarns with respect to weft ones (and vice-versa).
\end{itemize}

The continuum second-gradient model introduced in this paper must be seen as a reasonable engineering compromise between the easy finite element implementation of the proposed equations and the detail at which the complex behavior of the underlying microstructures is described. Despite its simplicity, the proposed model is able to 
\begin{itemize}
\item capture the main macroscopic S-shaped deformation mode of the considered unbalanced material
\item describe how this asymmetry in the macroscopic behavior is related to the differential local bending stiffness of the yarns at the mesoscopic
level 
\item include the presence of slipping of the yarns which has as a macroscopic counterpart an overall increment of the width within the specimen
\end{itemize}
Further studies should be mainly focused on
\begin{itemize}
\item a more precise interpretation and computation of the generalized second-gradient internal actions which may be present in the considered generalized
continua,
\item the development of suitable discrete models which are able to quantify the actual slippages of the yarns and relate it to the equivalent
elongations proposed here,
\item the setting up of more comprehensive experimental campaigns aimed at i) the precise estimation of experimental errors introduced during the tests, ii) the conception and development of independent tests for the validation of the proposed second-gradient model and iii) the study of eventual size effects in the considered unbalanced woven fabrics. 
\end{itemize}

\section*{Acknowledgements}

The authors thank CNRS for the PEPS project which assured financial support to research presented in this paper, the MESR for the Ph.D. scolarship of Gabriele Barbagallo and the ANR for the one of Ismael Azehaf.

\end{document}